\documentclass[showpacs,aps,graphicx,twocolumn]{revtex4}
\usepackage{amsmath}
\usepackage{amscd}
\usepackage{graphicx}

\begin{document}

\title{ Efficient quantum entanglement distribution over an arbitrary collective-noise channel\footnote{Published
in Phys. Rev. A \textbf{81}, 042332 (2010)}}

\author{Yu-Bo Sheng$^{1,2,3}$ and Fu-Guo Deng$^{1}$\footnote{
Corresponding author: fgdeng@bnu.edu.cn.} }
\address{$^1$Department of Physics, Beijing Normal University, Beijing 100875, China\\
$^2$College of Nuclear Science and Technology, Beijing Normal
University,
Beijing 100875, China\\
$^3$Key Laboratory of Beam Technology and Material Modification of
Ministry of Education, Beijing Normal University, Beijing 100875,
China}
\date{\today }

\begin{abstract}

We present an efficient quantum entanglement distribution over an
arbitrary collective-noise channel.  The basic idea in the present
scheme is that  two parties in quantum communication first transmit
the entangled states in the frequency degree of freedom which
suffers little from the  noise in an optical fiber. After the two
parties share the photon pairs, they add some operations and
equipments to transfer the frequency entanglement of pairs into the
polarization entanglement with the success probability of 100\%.
Finally, they can get maximally entangled polarization states with
polarization independent wavelength division multiplexers and
quantum frequency up-conversion which can erase distinguishability
for frequency. Compared with conventional entanglement purification
protocols, the present scheme works in a deterministic way in
principle. Surprisingly, the collective noise leads to an additional
advantage.
\end{abstract}
 \maketitle

\section{introduction}

Entanglement  between two distant locations is an essential resource
for quantum information and communication
\cite{computation1,computation2,book}. Many quantum information
processes cannot be realized perfectly without maximally entangled
states. For instance, quantum teleportation \cite{teleportation},
quantum  dense coding \cite{densecoding,super2}, and quantum-state
sharing \cite{QSTS} require entangled states to set up a quantum
channel between Alice and Bob, the two parties in quantum
communication. Also, Alice and Bob can exploit entangled photon
pairs to create a private key efficiently
\cite{Ekert91,BBM92,rmp,LongLiu,CORE,QKDLixh}, in particular in
long-distance quantum communication with quantum repeater
\cite{repeater1,repeater2,repeater3}. As photons are the best
physical systems for long-distance transmission of quantum states,
people always choose their entangled states in the polarization
degree of freedom to fulfill these tasks discussed previously.
However, during a practical transmission, the polarization degree of
freedom of photons is incident to be influenced by the thermal
fluctuation, vibration, and the imperfection of an optical fiber.
That is, they suffer from the channel noise inevitably whether they
are single photons or entangled photon pairs. Thus, various error
correction and error-rejection processes are proposed. For example,
with decoherent-free subspaces, Walton \emph{et al.} \cite{walton}
proposed a scheme for rejecting the errors introduced by a
collective noise. Quantum redundancy-code is also introduced to
solve this problem \cite{book}. For the faithful transmission of a
single-photon polarization state over a collective-noise channel,
Yamamoto et al. \cite{yamamoto} proposed an error-rejecting scheme
with an additional single photon in 2005. The success probability is
in principle 1/16 without two-qubit operations. Subsequently,
Kalamidas \cite{kalamidas} proposed two schemes to reject and
correct arbitrary qubit errors without additional particles, but
fast polarization modulators. In 2007, Li et al. \cite{xihan} also
proposed a faithful qubit transmission scheme against a collective
noise without ancillary qubits. Its success probability is 50\% with
only linear optical elements in a passive way. Also, they presented
another faithful single-qubit transmission scheme with a success
probability of 50\%  based on the frequency degree of freedom of
photons, resorting to an additional qubit \cite{xihanOC}.

For entangled quantum systems, there is another kind of processes
which can be used to decrease the influence arising from the noise,
named entanglement purification. For purifying a Werner state
\cite{werner},  Bennett \emph{et al.} \cite{Bennett1} proposed an
original entanglement purification protocol (EPP) based on quantum
controlled-NOT (CNOT) gates in 1996. Subsequently, several EPPs
based on similar quantum logic operations have been introduced. At
present, a perfect CNOT gate based on linear optical elements is
very difficult to   implement experimentally with  current
technology. In 2001, Pan \emph{et al.} \cite{Pan1} proposed an EPP
based on linear optics, without resorting to CNOT gates, which is
feasible in experiment. We also proposed an EPP based on cross-Kerr
nonlinearity \cite{shengpra}. However, entanglement purification is
essentially used to distill some high-fidelity entangled states from
less-entangled ones by sacrificing several qubits. In other words,
all conventional EPPs \cite{Bennett1,Pan1,shengpra,Simon} cannot get
perfect maximally entangled photon pairs by far as they work
probabilistically in principle. Thus, the faithful distribution of
maximally pure entangled states between two distant locations is
valuable for the realization of long-distance quantum communication.

The polarization entanglement of photon pairs is easily disturbed by
the noise in quantum channel, so it is not a good way to transmit
the polarization entanglement of photons directly over a noisy
channel. There are some other degrees of freedom of photons, which
suffer little from the channel noise over an optical fiber, such as
the spatial degree of freedom and the frequency degree of freedom of
photons. With present technology, the entanglement of photons in the
frequency degree of freedom is not difficult to be prepared with
spontaneous parametric down-conversion \cite{frequency1,frequency2}.
When a light propagates through an optically nonlinear medium with
second-order nonlinearity ($\chi^{2}$),  we can produce a pair of
photons in the "idler" and the "signal" modes. Also the conservation
in energy and momentum  can give rise to entanglement in various
degrees of freedom, such as polarization entanglement, time-energy
entanglement, and position-momentum entanglement.

In this paper, we present an  efficient entanglement distribution
scheme over an arbitrary collective-noise channel. The basic idea of
the present scheme is that the two parties, say Alice and Bob, first
transmit an entangled state in the frequency degree of freedom which
suffers little from the channel noise in an optical fiber. After
Alice and Bob share an entangled photon pair, they add some
operations and equipments to transfer the frequency entanglement of
the photon pair into the polarization entanglement with a success
probability of 100\%. Compared with conventional entanglement
purification, this scheme does not require quantum resources
largely. Our protocol has several advantages. First,  the noise
channel can be an arbitrarily collective one. Second, the two
parties can get a perfect maximally entangled state in polarization
in principle. In a practical transmission, Alice and Bob can also
obtain perfect maximally entangled states in polarization by
controlling the distances between the entangled source and the
users. Moreover, this protocol can be generalized to distribute a
multipartite entangled quantum system and can be easily realized in
current experimental conditions.

\section{Efficient quantum entanglement distribution of two-qubit systems}

Cross-Kerr nonlinearity is a powerful tool for us to construct
nondestructive quantum nondemoliton detectors (QND)
\cite{QND1,QND2}. The cross-Kerr nonlinearity has been used to
prepare  CONT gates \cite{QND1} and complete a local Bell-state
analysis \cite{QND2}. Also it can be used to fulfill the quantum
entanglement purification and entanglement concentration protocols
\cite{shengpra,shengpra2}. The Hamiltonian of the cross-Kerr
nonlinearity is $H_{ck}=\hbar\chi a^{+}_{s}a_{s}a^{+}_{p}a_{p}$
\cite{QND1,QND2}. Here $a^{+}_{s}$ and $a^{+}_{p}$ are the creation
operations and $a_{s}$ and $a_{p}$ are the destruction operations.
Suppose a signal state
$|\Psi\rangle_s=c_{0}|0\rangle_{s}+c_{1}|1\rangle_{s}$
($|0\rangle_{s}$ and $|1\rangle_{s}$ denote that there are no photon
and one photon, respectively, in this state) and a coherent probe
beam in the state $|\alpha\rangle$ couple with a cross-Kerr
nonlinearity medium, the whole system evolves as
\begin{eqnarray}
U_{ck}|\Psi\rangle_{s}|\alpha\rangle_{p}&=&
e^{iH_{ck}t/\hbar}[c_{0}|0\rangle_{s}+c_{1}
|1\rangle_{s}]|\alpha\rangle_{p} \nonumber\\
&=& c_{0}|0\rangle_{s}|\alpha\rangle_{p}+c_{1}|1\rangle_{s}|
\alpha e^{i\theta}\rangle_{p},
\end{eqnarray}
where $\theta=\chi t$ and $t$ is the interaction time. The coherent
beam picks up a phase shift $\theta$ directly proportional to the
number of the photons in the Fock state $|\Psi\rangle_s$, which can
be read out with  a general homodyne-heterodyne measurement. So one
can exactly check the number of photons in the Fock state but not
destroy them.

Now let us explain the principle of our entanglement distribution
protocol over an arbitrary collective-noise channel. We suppose that
the center, say Carl prepares an entangled photon pair $ab$ in the
following state:
\begin{eqnarray}
|\Psi\rangle_{ab}=\frac{1}{\sqrt{2}}|H\rangle_{a}|H\rangle_{b}
(|\omega_{1}\rangle|\omega_{2}\rangle
+|\omega_{2}\rangle|\omega_{1}\rangle).
\end{eqnarray}
Here we denote the state of a horizontally polarized photon by
$|H\rangle$ and the state of a vertically polarized photon by
$|V\rangle$. $|\omega_{1}\rangle|\omega_{2}\rangle$ and
$|\omega_{2}\rangle|\omega_{1}\rangle$ are two different frequency
modes of the two photons. The subscripts $a$ and $b$ mean that the
two photons  are distributed to Alice and Bob, respectively. Suppose
the collective noises in the two channels have the same form but
different noise parameters which alter with time in principle, i.e.,
\begin{eqnarray}
|H\rangle_{a} &\rightarrow & \alpha|H\rangle+\beta|V\rangle,\nonumber\\
|H\rangle_{b} &\rightarrow & \delta|H\rangle+\gamma|V\rangle,
\end{eqnarray}
where
\begin{eqnarray}
|\alpha|^{2}+|\beta|^{2}=1,\;\;\;\;\;
|\delta|^{2}+|\gamma|^{2}=1.
\end{eqnarray}

The photon pair in the input modes of the channels will suffer from
two collective noises, shown in Fig. 1, that is, the evolution of
its state through the noisy channels can be written as:
\begin{eqnarray}
|\Psi\rangle_{ab} &=&
\frac{1}{\sqrt{2}}(|H\rangle_{\omega_{1}}|H\rangle_{\omega_{2}}
+|H\rangle_{\omega_{2}}|H\rangle_{\omega_{1}}) \xrightarrow{ noises} \nonumber\\
|\Psi\rangle'_{ab}& =&
\frac{1}{\sqrt{2}}[(\alpha|H\rangle_{\omega_{1}}+\beta|V\rangle_{\omega_{1}})(\delta|H\rangle_{\omega_{2}}
+\gamma|V\rangle_{\omega_{2}})\nonumber\\
&&\;\;\;\; +
(\alpha|H\rangle_{\omega_{2}}+\beta|V\rangle_{\omega_{2}})(\delta|H\rangle_{\omega_{1}}
+\gamma|V\rangle_{\omega_{1}})]\nonumber\\
&=&\frac{1}{\sqrt{2}}[\alpha\delta|H\rangle_{\omega_{1}}|H\rangle_{\omega_{2}}
+\alpha\gamma|H\rangle_{\omega_{1}}|V\rangle_{\omega_{2}}\nonumber\\
&&\;\;\;\; +\beta\delta|V\rangle_{\omega_{1}}|H\rangle_{\omega_{2}}
+\beta\gamma|V\rangle_{\omega_{1}}|V\rangle_{\omega_{2}}\nonumber\\
&&\;\;\;\; +\alpha\delta|H\rangle_{\omega_{2}}|H\rangle_{\omega_{1}}
+\alpha\gamma|H\rangle_{\omega_{2}}|V\rangle_{\omega_{1}}\nonumber\\
&&\;\;\;\; +\beta\delta|V\rangle_{\omega_{2}}|H\rangle_{\omega_{1}}
+\beta\gamma|V\rangle_{\omega_{2}}|V\rangle_{\omega_{1}}].
\end{eqnarray}
After the noisy channel, the photon $a$ ($b$) will pass through a
polarization beam splitter (PBS) which transmits the horizontal
polarization mode $\vert H \rangle$ and reflects the vertical
polarization mode $\vert V \rangle$. If Alice and Bob combine their
photons and their coherent probe beams ($\vert \alpha\rangle_A$ and
$\vert \alpha\rangle_B$) with cross-Kerr nonlinearity media (shown
in Fig.1.), the state of whole quantum system becomes
\begin{eqnarray}
\rightarrow
&&\frac{1}{\sqrt{2}}[\alpha\delta(|H\rangle_{\omega_{1}}|H\rangle_{\omega_{2}}
+|H\rangle_{\omega_{2}}|H\rangle_{\omega_{1}})|\alpha
e^{i\theta}\rangle_{A}|\alpha e^{i\theta}\rangle_{B}\nonumber\\
&&\;\;\;\;
+\alpha\gamma(|H\rangle_{\omega_{1}}|V\rangle_{\omega_{2}}+|H\rangle_{\omega_{2}}|V\rangle_{\omega_{1}})|\alpha
e^{i\theta}\rangle_{A}|\alpha e^{i\theta'}\rangle_{B}\nonumber\\
&&\;\;\;\;
+\beta\delta(|V\rangle_{\omega_{1}}|H\rangle_{\omega_{2}}+|V\rangle_{\omega_{2}}|H\rangle_{\omega_{1}})|\alpha
e^{i\theta'}\rangle_{A}|\alpha e^{i\theta}\rangle_{B}\nonumber\\
&&\;\;\;\;
+\beta\gamma(|V\rangle_{\omega_{1}}|V\rangle_{\omega_{2}}+|V\rangle_{\omega_{2}}|V\rangle_{\omega_{1}})|\alpha
e^{i\theta'}\rangle_{A}|\alpha e^{i\theta'}\rangle_{B}].\nonumber
\end{eqnarray}
Here $|\alpha e^{i\theta}\rangle_{A}$ means that the coherent probe
beam in Alice's hand picks up a phase shift $\theta$. The other
terms are analogical with it.

\begin{figure}[!h]
\begin{center}
\includegraphics[width=8cm,angle=0]{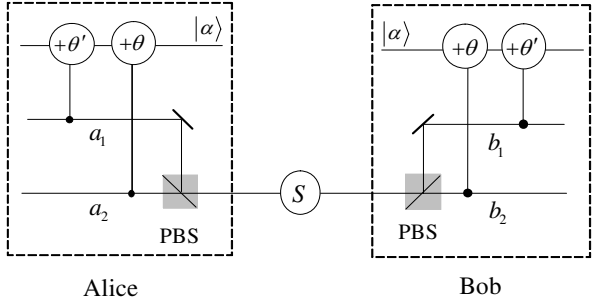}
\caption{(Color online)Schematic drawing of quantum entanglement
distribution over a collective-noise channel with QND. The
entanglement source (S) produces two entangled photons which are
transmitted to  two parties in quantum communication, say Alice and
Bob, respectively. The two photons suffer from the noise during the
transmission. PBS presents a polarization beam splitter. Alice and
Bob can check the phase shifts of their coherent beams to judge
which state they obtain in a deterministic way. $+\theta$ and
$+\theta'$ represent two cross-Kerr nonlinear media with the phase
shifts $+\theta$ and $+\theta'$, respectively. }
\end{center}
\end{figure}

After $X$ homodyne measurements on their coherent beams
independently, Alice and Bob will get some different phase shifts
and the photon pair will appear at some different output modes. In
detail, if Alice and Bob have the same phase shift $\theta$,  the
photon pair $ab$ collapses to the state $\vert
\phi_1\rangle_{ab}=\frac{1}{\sqrt{2}}(|H\rangle_{\omega_{1}}|H\rangle_{\omega_{2}}
+|H\rangle_{\omega_{2}}|H\rangle_{\omega_{1}})_{ab}$ and they will
appear at the lower output modes $a_{2}b_{2}$. If Alice and Bob have
the same phase shift $\theta'$,  the photon pair $ab$ collapses to
the state $\vert
\phi_2\rangle_{ab}=\frac{1}{\sqrt{2}}(|V\rangle_{\omega_{1}}|V\rangle_{\omega_{2}}
+|V\rangle_{\omega_{2}}|V\rangle_{\omega_{1}})_{ab}$ and they will
appear at the upper output modes $a_{1}b_{1}$. The state $\vert
\phi_3\rangle_{ab}=\frac{1}{\sqrt{2}}(|H\rangle_{\omega_{1}}|V\rangle_{\omega_{2}}
+|H\rangle_{\omega_{2}}|V\rangle_{\omega_{1}})_{ab}$ will be in the
output modes $a_{2}$ and $b_{1}$, which leads the phase shift
$\theta$ in Alice and $\theta'$ in Bob. $\vert
\phi_4\rangle_{ab}=\frac{1}{\sqrt{2}}(|V\rangle_{\omega_{1}}|H\rangle_{\omega_{2}}
+|V\rangle_{\omega_{2}}|H\rangle_{\omega_{1}})_{ab}$ leads the phase
shift $\theta'$ in Alice and $\theta$ in Bob, and the photon pair
will appear at the output modes $a_1b_2$. That is, with $X$ homodyne
measurements Alice and Bob can distinguish the four entangled states
$\vert \phi_i\rangle_{ab}$ ($i=1,2,3,4$).

\begin{figure}[!h]
\begin{center}
\includegraphics[width=8cm,angle=0]{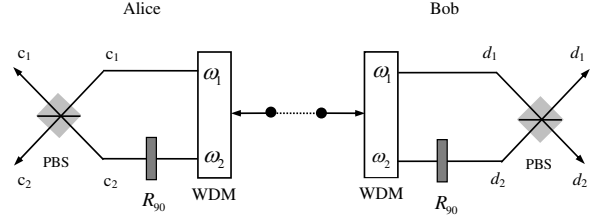}
\caption{Schematic drawing of converting  frequency entanglements to
polarization entanglements. WDM  guides photons to different spatial
modes according to their different frequencies. $R_{90}$ represents
a rotation by 90$^\circ$, which acts as a bit-flip operation and
completes the transformation between the horizontal polarization $H$
and the vertical polarization $V$.}
\end{center}
\end{figure}

The second step of our entanglement distribution protocol is to
convert the frequency-entangled states $\vert \phi_i\rangle_{ab}$ to
polarization-entangled ones. We take
$\vert\phi_1\rangle_{ab}=\frac{1}{\sqrt{2}}(|H\rangle_{\omega_{1}}|H\rangle_{\omega_{2}}
+|H\rangle_{\omega_{2}}|H\rangle_{\omega_{1}})_{ab}$ as an example
to describe the principle of this step, shown in Fig.2. WDM
represents a polarization independent wavelength division
multiplexer. It can be used to guide  photons to  different spatial
modes according to their frequencies. For Alice (Bob), the photons
with the frequencies $\omega_1$ and $\omega_2$ will be guided to the
spatial modes $c_1$ ($d_1$) and $c_2$ ($d_2$), respectively. Two
wave plates $R_{90^{\circ}}$ are used to rotate the horizontal
polarization $H$ and the vertical polarization $V$ by $90^{\circ}$.
That is, they complete the transformation $|H\rangle \rightarrow
|V\rangle$ and $|V\rangle \rightarrow |H\rangle$. This task can be
accomplished with a half-wave plate whose orientation is
45$^{\circ}$. After the photon pair $ab$ is coupled by the two PBSs,
its state  becomes
\begin{eqnarray}
\rightarrow \vert
\phi'_1\rangle_{ab}=\frac{1}{\sqrt{2}}(|H\rangle_{\omega_{1}}|V\rangle_{\omega_{2}}
+|V\rangle_{\omega_{2}}|H\rangle_{\omega_{1}})
\end{eqnarray}
and will be in the output modes $c_{2}$ and $d_{2}$. Following the
similar way, Alice and Bob can obtain the other three entangled
states $\vert \phi'_2\rangle$, $\vert \phi'_3\rangle$, and $\vert
\phi'_4\rangle$
 in the output modes
$c_{1}d_{1}$, $c_2d_1$, and $c_1d_2$, respectively. Here
\begin{eqnarray}
\vert \phi'_2\rangle &=&
\frac{1}{\sqrt{2}}(|V\rangle_{\omega_{1}}|H\rangle_{\omega_{2}}
+|H\rangle_{\omega_{2}}|V\rangle_{\omega_{1}}),\\
\vert \phi'_3\rangle &=&
\frac{1}{\sqrt{2}}(|H\rangle_{\omega_{1}}|H\rangle_{\omega_{2}}
+|V\rangle_{\omega_{2}}|V\rangle_{\omega_{1}}),\\
\vert \phi'_4\rangle &=&
\frac{1}{\sqrt{2}}(|V\rangle_{\omega_{1}}|V\rangle_{\omega_{2}}
+|H\rangle_{\omega_{2}}|H\rangle_{\omega_{1}}).
\end{eqnarray}

Each of the four states $\{\vert \phi'_1\rangle, \vert
\phi'_2\rangle, \vert \phi'_3\rangle, \vert \phi'_4\rangle\}$ is a
maximally entangled one in both polarization and frequency degrees
of freedom. Alice and Bob can erase the distinguishability for the
frequency of their photons with the help of quantum frequency
up-conversion \cite{erase-frequency} and turn them into a standard
Bell state $\vert \phi^+\rangle_{ab}=\frac{1}{\sqrt{2}}(|H\rangle
|H\rangle + |V\rangle|V\rangle)$ with local unitary operations.
Moreover, the success probability of this entanglement distribution
scheme is in principle 100\% over an arbitrary collective-noise
channel as it is independent of the noise parameters $\alpha$,
$\beta$, $\delta$, and $\gamma$, which is different from
single-photon error-rejecting protocols
\cite{walton,yamamoto,kalamidas,xihan,xihanOC}.

\section{Efficient quantum entanglement distribution of multiqubit systems}

This scheme can be generalized for distribution of $n$-qubit system
($n>2$) in a Greenberger-Horne-Zeilinger (GHZ) state over an
arbitrary collective-noise channel. Let us use the distribution of a
four-qubit system as an example  to describe its principle. The
other cases are similar to it with or without a little of
modification.

Suppose that the initial state of a four-qubit system is
\begin{eqnarray}
|\Phi_4\rangle_{ABCD}=\frac{1}{\sqrt{2}}\vert 0000\rangle(\vert
\omega_{1}\omega_{2}\omega_{1}\omega_{2}\rangle + \vert
\omega_{2}\omega_{1}\omega_{2}\omega_{1}\rangle)_{ABCD}\nonumber
\end{eqnarray}
and the collective noises in the four channels have the same form
but different noise parameters which alter with time $t$ in
principle, i.e.,
\begin{eqnarray}
|0\rangle_{i}\rightarrow  \beta^i_0|0\rangle + \beta^i_1|1\rangle,
\end{eqnarray}
where $i=A,B,C,D$ represent the four photons which are sent to
Alice, Bob, Charlie, and Daniel, respectively. Here $\vert 0\rangle
\equiv \vert H\rangle$ and $\vert 1\rangle \equiv \vert V\rangle$.
After passing through the noisy channels, the four-qubit system
evolves  as
\begin{eqnarray}
|\Phi_4\rangle_{ABCD}& & \xrightarrow{{\tiny noises}} \nonumber\\
|\Phi_4\rangle'_{ABCD} &=&\frac{1}{\sqrt{2}}\left( \sum_{jklm}
\beta^A_j\beta^B_k\beta^C_l\beta^D_m\vert j\rangle_A\vert k\rangle_B
\vert l\rangle_C \vert m\rangle_D\right)\nonumber\\
&& \;\;\;\;\;\; \cdot \left(\vert
\omega_{1}\omega_{2}\omega_{1}\omega_{2}\rangle + \vert
\omega_{2}\omega_{1}\omega_{2}\omega_{1}\rangle\right)_{ABCD},\nonumber\\
\end{eqnarray}
where $j,k,l,m \in\{0,1\}$. Similar to Fig.1, Alice, Bob, Charlie,
and Daniel use their QNDs to check the polarization states of their
photons. That is, if one obtains the phase shift of his coherent
beam $\theta$, his photon is in the polarization state $\vert
0\rangle=\vert H\rangle$; otherwise, the photon is in $\vert
1\rangle=\vert V\rangle$. With their outcomes of their  $X$ homodyne
measurements and some local unitary operations, the four users can
obtain the state $|\Phi_4\rangle''_{ABCD}=\frac{1}{\sqrt{2}}\vert
0000\rangle(\vert \omega_{1}\omega_{2}\omega_{1}\omega_{2}\rangle +
\vert \omega_{2}\omega_{1}\omega_{2}\omega_{1}\rangle)_{ABCD}$. With
the setups similar to Fig. 2, the four users can obtain the
entangled state in polarization
$|\Psi_4\rangle''_{ABCD}=\frac{1}{\sqrt{2}}(\vert
H_{\omega_{1}}V_{\omega_{2}}H_{\omega_{1}}V_{\omega_{2}\rangle} +
V_{\vert
\omega_{2}}H_{\omega_{1}}V_{\omega_{2}}H_{\omega_{1}}\rangle)_{ABCD}$.
Alice, Bob, Charlie, and Daniel can erase the distinguishability for
the frequencies of their photons with the help of quantum frequency
up-conversion \cite{erase-frequency} and turn their system into a
GHZ state $\vert \Psi\rangle'_{ABCD}=\frac{1}{\sqrt{2}}(|H\rangle
|V\rangle|H\rangle|V\rangle +
|V\rangle|H\rangle|V\rangle|H\rangle)_{ABCD}$. With two bit-flip
operations on the photons $B$ and $D$, respectively, they will
obtain a standard GHZ state $\vert
\Psi\rangle_{ABCD}=\frac{1}{\sqrt{2}}(|H\rangle
|H\rangle|H\rangle|H\rangle +
|V\rangle|V\rangle|V\rangle|V\rangle)_{ABCD}$.

\section{discussion and summary}

We have discussed our quantum entanglement distribution scheme in
the case that the frequency degree of freedom of photon pairs is
insensitive to channel noise.  The previous experiments showed that
the polarization entanglement is quite unsuitable for transmission
over distances of more than a few kilometers in an optical fiber
\cite{rmp}. For example, Naik \emph{et al.} demonstrated the Ekert
protocol \cite{Ekert91} by only a few meters \cite{experiment1,rmp}.
Also, they observed the quantum bit error rate (QBER) increase to
33\% in the experiment implementation of the six-state protocol
\cite{sixstate1,sixstate2}. For frequency coding
\cite{experiment2,experiment5,experiment3,experiment4,frequencystable1,frequencystable2},
for example, the Besancon group performed a key distribution over a
20-km single-mode optical-fiber spool. They recorded a QBER$_{opt}$
contribution of approximately 4\%, and estimated that 2\% could be
attributed to the transmission of the central frequency by the
Fabry-Perot cavity \cite{experiment5}. That is, on one hand, the
channel noise less affects the entanglement in the frequency degree
of freedom. On the other hand, the optical fibers used to transmit
photons will introduce a relative phase on the entanglement as there
are two different frequencies in each photon. That is, the entangled
state in the frequency degree of freedom will become
$\frac{1}{\sqrt{2}}(|\omega_{1}\omega_{2}\rangle + e ^{i\Delta
\phi_{f}} |\omega_{2}\omega_{1}\rangle)$ after the two photons $a$
and $b$ are sent to Alice and Bob, respectively. Here $\Delta
\phi_{f}\equiv \frac{1}{v}[(\omega_{2}-\omega_{1})L_A +
(\omega_{1}-\omega_{2})L_B]$. $v$ and $L_A$ ($L_B$) represent the
velocity of photons in an optical fiber and the distance between the
entangled source and Alice (Bob), respectively. When $L_A=L_B$,
$\Delta \phi_{f}=0$. That is, Alice and Bob can obtain a perfect
entangled state in the frequency degree of freedom after their
transmission if they can control their distances between them and
the entangled source. Also, Alice and Bob can compensate the
relative phase $\Delta \phi_{f}$ after their transmission if $L_A
\neq L_B$, as $\Delta \phi_{f}$ is in general invariable and can be
detected. In this case, the relative phase $\Delta \phi_{f}$  in
frequency will be transferred into the entanglement in polarization.
That is, Alice and Bob will obtain the maximally entangled state in
the polarization degree of freedom with the form $\vert
\phi'^+\rangle_{ab}=\frac{1}{\sqrt{2}}(|H\rangle |H\rangle +
e^{i\Delta \phi_{f}} |V\rangle|V\rangle)$. With some unitary
operations by wave plates, they will obtain the standard Bell state
$\vert \phi^+\rangle_{ab}=\frac{1}{\sqrt{2}}(|H\rangle |H\rangle +
|V\rangle|V\rangle)$.

Let us compare this distribution scheme with conventional
entanglement purification protocols
\cite{Bennett1,Pan1,shengpra,Simon}. In the latter,  the two parties
transmit the entangled photon pairs in the polarization degree of
freedom directly over a noisy channel. The photon pair transmitted
suffers from the channel noise and its state becomes a mixed
entangled one. In Ref. \cite{Pan1}, the two sources produce two
pairs of entangled photons and one photon from each pair is
distributed to Alice and the other one to Bob. The two photons in
each side overlap at a PBS. By selecting the four-mode instances,
Alice and Bob can thus obtain a subset of  high-fidelity entangled
photon pairs. In order to get the entangled states with a higher
fidelity, Alice and Bob should repeat this protocol and consume more
less-entangled states. Ref. \cite{Simon} presented a more practical
polarization entanglement purification using spatial entanglement.
In their protocol, the parametric down-conversion source produces an
entangled photon pair in both polarization and spatial degrees of
freedom. By selecting those events where photons are both in the
upper mode or in the lower mode, the two parties can purify the
bit-flip error. However, both of these two protocols can not get
perfect maximally entangled pairs and they can only improve the
fidelity of an ensemble in a mixed entangled state by consuming the
quantum resource exponentially. The present scheme exploit the
entanglement in the frequency degree of freedom to create the
entanglement in the  polarization degree of freedom perfectly. After
the homodyne detectors, the entanglement in  the frequency degree of
freedom does not degrade in principle, which makes the present
scheme work in a deterministic way. This is different from the
entanglement purification protocols as the polarization entanglement
is degraded in the  noise channel in the latter. So the yield of
entanglement purification protocols is far lower than the present
scheme. This result is kept for the case with entanglement
concentration protocols \cite{zhao1,shengpra2} as the latter also
needs to sacrifice the less-entangled states largely to obtain a
maximally entangled one. Compared with the deterministic
entanglement purification protocol \cite{hep}, the present scheme
requires less entanglement resource as the former resorts to
hyperentanglement in three degrees of freedom (such as polarization,
spatial mode, and frequency) while the latter only resorts to the
entanglement in the frequency degree of freedom. Compared with the
faithful distribution of single-qubit scheme with linear optics
\cite{xihan}, the success probability of the present scheme is 100\%
while that of single-photon error-rejecting protocol \cite{xihan} is
only 50\%. That is, the present scheme may more practical for
distribution of entanglement in quantum communication with the
development of techniques.

We should point out that cross-Kerr effect is yet not easy to
implement in current experiment. The largest natural cross-Kerr
nonlinearities are extremely weak
($\chi^{(3)}\approx10^{-22}m^{2}V^{-2}$) \cite{QND3}. In Ref.
\cite{QND4}, Kok\emph{ et al.} showed that operating in the optical
single-photon regime, the Kerr phase shift is only
$\tau\approx10^{-18}$. With electromagnetically induced transparent
materials, cross-Kerr nonlinearities of $\tau\approx10^{-5}$ can be
obtained. The weak cross-Kerr nonlinearity will make the phase shift
$\theta$ and $\theta'$ of the coherent state became extremely small,
which will be hard to detect. That is to say, using homodyne
detector, it is difficult to determine the phase shift due to the
impossible discrimination of two overlapping coherent states, which
will decrease the success probability of the present scheme to 1/4
at worst. In 2003, Hofmann \emph{et al.} showed that a phase shift
of $\pi$ can be achieved with a single two-level atom in a one-sided
cavity \cite{QND5}. In 2010, Wittmann \emph{et al.} investigated
quantum measurement strategies capable of discriminating two
coherent states using a homodyne detector and a photon number
resolving (PNR) detector \cite{discrimination}. In order to lower
the error probability, the postselection strategy is applied to the
measurement data of homodyne detector as well as a PNR detector.
They indicated that the performance of the new displacement
controlled PNR is better than homodyne receiver.

In summary, we have presented an efficient entanglement distribution
scheme over an arbitrary collective-noise channel. Compared with
conventional entanglement purification protocols
\cite{walton,yamamoto,kalamidas,xihan,xihanOC}, the present scheme
does not consume a great deal of less-entangled resources and it
works in a determinate way. In essence, it is the entanglement
transformation between two different degrees of freedom of photons.
We exploit the feature that the frequency of photons suffers little
from the channel noise to generate the entanglement in the
polarization degree of freedom. If other degrees of freedom are
robust to the channel noise, they also can be used to implement our
protocol, and the frequency degree of freedom is not unique. We
believe that the present scheme for the distribution of entangled
states in the polarization degree of freedom may be a vital
ingredient in the realization of long-distance quantum communication
in the future.

\section*{ACKNOWLEDGEMENTS}

This work is supported by the National Natural Science Foundation of
China under Grant No. 10974020, A Foundation for the Author of
National Excellent Doctoral Dissertation of P. R. China under Grant
No. 200723, and  Beijing Natural Science Foundation under Grant No.
1082008.


\begin{thebibliography}{99}

\bibitem{computation1} D. P. Divincenzo, Science  \textbf{270}, 255 (1995).

\bibitem{computation2} C. H. Bennett and D. P. Divincenzo,
Nature (London) \textbf{404}, 247 (2000).

\bibitem{book} M. A. Nielsen and I. L. Chuang, \emph{Quantum Computation and
Quantum Information} (Cambridge University Press, Cambridge,
 2000).


\bibitem{teleportation} C. H. Bennett, G. Brassard, C. Crepeau, R. Jozsa, A. Peres,
and W. K. Wootters, Phys. Rev. Lett.
\textbf{70}, 1895 (1993).


\bibitem{densecoding} C. H. Bennett and S. J. Wiesner, Phys. Rev. Lett. \textbf{69}, 2881 (1992).

\bibitem{super2} X. S. Liu, G. L. Long, D. M. Tong, and L. Feng, Phys. Rev. A \textbf{65},
022304 (2002).

\bibitem{QSTS} F. G. Deng, X. H. Li, C. Y. Li, P. Zhou, and H. Y.
Zhou, Phys. Rev. A \textbf{72}, 044301 (2005).


\bibitem{Ekert91} A. K. Ekert,  Phys. Rev. Lett. \textbf{67},
  661 (1991).

\bibitem{BBM92} C. H. Bennett, G. Brassard, and N. D. Mermin,
 Phys. Rev. Lett. \textbf{68},  557 (1992).

\bibitem{rmp} N. Gisin, G. Ribordy, W. Tittel, and H. Zbinden
Rev. Mod. Phys.  \textbf{74}, 145  (2002).



\bibitem{LongLiu} G. L. Long and X. S. Liu, Phys. Rev. A  \textbf{65}, 032302 (2002).

\bibitem{CORE} F. G. Deng and G. L. Long, Phys. Rev. A  \textbf{68},
042315 (2003).


\bibitem{QKDLixh} X. H. Li, F. G. Deng, and H. Y. Zhou, Phys. Rev. A
\textbf{78}, 022321 (2008).

\bibitem{repeater1} L. M. Duan, M. D. Lukin, J. I. Cirac, and P. Zoller,
Nature (London) \textbf{414}, 22 (2001).

\bibitem{repeater2} C. Simon, H. de Riedmatten, M. Afzelius, N.
Sangouard, H. Zbinden, and N. Gisin, Phys. Rev. Lett. \textbf{98},
190503 (2007).

\bibitem{repeater3} N. Sangouard, C. Simon, J. Minar, H. Zbinden, H. deRiedmatten, and N.
Gisin, Phys. Rev. A \textbf{76}, 050301(R) (2007).




\bibitem{walton} Z. D. Walton, A. F. Abouraddy, A. V. Sergienko, B. E. A. Saleh,
and M. C. Teich, Phys. Rev. Lett. \textbf{91}, 087901 (2003).


\bibitem{yamamoto} T. Yamamoto, J. Shimamura, S. K. \"Ozdemir, M.
Koashi, and N. Imoto, Phys. Rev. Lett. \textbf{95}, 040503 (2005).


\bibitem{kalamidas} D. Kalamidas, Phys. Lett. A  \textbf{343}, 331 (2005).


\bibitem{xihan} X. H. Li, F. G. Deng, and H. Y. Zhou, Appl. Phys.
Lett. \textbf{91}, 144101 (2007).


\bibitem{xihanOC} X. H. Li, B. K. Zhao, Y. B. Sheng, F. G. Deng, and
H. Y. Zhou, Opt. Commun. \textbf{282}, 4025 (2009).


\bibitem{werner} R. F. Werner, Phys. Rev. A \textbf{40}, 4277 (1989).



\bibitem{Bennett1} C. H. Bennett, G. Brassard, S. Popescu, B. Schumacher,
J. A. Smolin and W. K. Wootters, Phys. Rev. Lett \textbf{76}, 722 (1996).


\bibitem{Pan1} J. W. Pan, C. Simon, and A. Zellinger,  Nature (London)
\textbf{410}, 1067 (2001)



\bibitem{shengpra} Y. B. Sheng, F. G. Deng, and H. Y. Zhou, Phys.
Rev. A. \textbf{77}, 042308 (2008).


\bibitem{Simon} C. Simon and J. W. Pan,  Phys. Rev. Lett.
\textbf{89}, 257901 (2002).


\bibitem{frequency1} A. Yabushita  and T. Kobayashi, Phys.
Rev. A. \textbf{69}, 013806 (2004).

\bibitem{frequency2} A. Yabushita  and T. Kobayashi, J.  Appl.
Phys. \textbf{99}, 060301 (2006).

\bibitem{QND1} K. Nemoto and W. J. Munro,  Phys. Rev. Lett. \textbf{93}, 250502 (2004).

\bibitem{QND2} S. D. Barrett, P. Kok, K. Nemoto,  R. G. Beausoleil,  W. J. Munro, and T. P. Spiller,
 Phys. Rev. A \textbf{71},
 060302(R) (2005).



\bibitem{shengpra2} Y. B. Sheng, F. G. Deng, and H. Y. Zhou, Phys.
Rev. A. \textbf{77}, 062325 (2008).


\bibitem{erase-frequency} H. Takesue,   Phys. Rev. Lett.
\textbf{101}, 173901 (2008).






\bibitem{experiment1}D. S. Naik, C. G. Peterson, A. G. White, A. J. Berglund, and P. G. Kwiat,
Phys. Rev. Lett. \textbf{84}, 4733 (2000).




\bibitem{sixstate1} D. Bruss, Phys. Rev. Lett. \textbf{81}, 3018
(1998).

\bibitem{sixstate2} H. Bechmann-Pasquinucci and N. Gisin, Phys. Rev. A \textbf{59}, 4238
(1999).




\bibitem{frequencystable1} T. Zhang,  Z. Q.  Yin,  T. F. Han, and G. C. Guo,
Opt. Commun. \textbf{281}, 4800 (2008).



\bibitem{frequencystable2} M. Bloch, S. W. McLaughlin,  J. M. Merolla, and F. Patois,  Opt. Lett. \textbf{32},
301 (2007).

\bibitem{experiment2} B. Huttner, N. Imoto, N. Gisin, and T. Mor,,
Phys. Rev. A \textbf{51}, 1863 (1995).

\bibitem{experiment3} P. C. Sun, Y. Mazurenko, and Y. Fainman,
 Opt. Lett. \textbf{20}, 1062 (1995).

 \bibitem{experiment4} Y. Mazurenko, R.Giust, and J. P. Goedgebuer,,
Opt. Commun. \textbf{133}, 87 (1997).



\bibitem{experiment5}J. M. Merolla, Y. Mazurenko, J. P. Goedgebuer, and W. T. Rhodes,
 Phys. Rev. Lett. \textbf{82}, 1656 (1999).



\bibitem{zhao1} Z. Zhao, J. W. Pan, and M. S. Zhan, Phys. Rev. A \textbf{64}, 014301 (2001).


\bibitem{hep} Y. B. Sheng and F. G. Deng, Phys. Rev. A \textbf{81}, 032307 (2010).

\bibitem{QND3} P. Kok, W. J. Munro, K. Nemoto, T. C. Ralph, J. P. Dowling, and G. J. Milburn, Rev. Mod. Phys.  \textbf{79}, 135 (2007).

\bibitem{QND4} P. Kok, H. Lee, and J. P. Dowling, Phys. Rev. A  \textbf{66}, 063814 (2002).

\bibitem{QND5} H. F. Hofmann, K. Kojima, S. Takeuchi, and K. Sasaki,
J. Opt. B: Quantum Semiclassical Opt.  \textbf{5}, 218 (2003).

\bibitem{discrimination} C. Wittmann, U. L. Andersen, M. Takeoka,
and G. Leuchs, e-print arXiv: 1002.0232.

\end{thebibliography}
\end{document}